\RequirePackage{fix-cm}

\documentclass[twoside,11pt]{article}
\usepackage{journal, theapa, rawfonts}

\usepackage{amssymb}
\usepackage{amsmath}
\usepackage{graphicx}
\usepackage{subfig}
\usepackage{multirow}
\usepackage{latexsym}
\usepackage{rotating}
\usepackage{wrapfig}
\usepackage{floatflt}
\usepackage{arydshln}
\usepackage{graphics}
\usepackage{mathtools}
\usepackage{natbib}
\bibpunct{(}{)}{;}{a}{,}{,}
\usepackage[colorlinks=false]{hyperref}

\usepackage{url}
\usepackage{xcolor}
\definecolor{newcolor}{rgb}{.8,.349,.1}

\makeatletter
\newcommand{\onset}[2]{%
  \;{\mathop{#1}\limits_{\vbox to 5\ex@{\kern-\tw@\ex@
   \hbox{\scriptsize #2}\vss}}}\;}
\makeatother

\begin{document}

\title{A new simple and effective measure for bag-of-word \\ inter-document similarity measurement}
%\title{A new simple and effective inter-document similarity measure in the bag-of-words vector space model}
%\title{A new simple and effective inter-document similarity measure to overcome the shortcomings of term weighting used by existing measures in the bag-of-words vector space model}

\author{\name Sunil Aryal$^{1}$ \email sunil.aryal@deakin.edu.au \\
        \name Kai Ming Ting$^2$ \email kaiming.ting@federation.edu.au \\
        \name Takashi Washio$^3$ \email washio@ar.sanken.osaka-u.ac.jp \\
        \name Gholamreza Haffari$^4$ \email gholamreza.haffari@monash.edu \\
        \addr $^1$Deakin University, Australia \\   
        \addr $^2$Federation University, Australia \\ 
        \addr $^3$Osaka University, Japan \\
        \addr $^4$Monash University, Australia}

\date{}

\maketitle

%==== Abstract =====

\begin{abstract}
To measure the similarity of two documents in the bag-of-words (BoW) vector representation, different term weighting schemes are used to improve the performance of cosine similarity---the most widely used inter-document similarity measure in text mining. In this paper, we identify the shortcomings of the underlying assumptions of term weighting in the inter-document similarity measurement task; and provide a more fit-to-the-purpose alternative. Based on this new assumption, we introduce a new simple but effective similarity measure which does not require explicit term weighting. The proposed measure employs a more nuanced probabilistic approach than those used in term weighting to measure the similarity of two documents w.r.t each term occurring in the two documents. Our empirical comparison with the existing similarity measures using different term weighting schemes shows that the new measure produces (i) better results in the binary BoW representation; and (ii) competitive and more consistent results in the term-frequency-based BoW representation.
\end{abstract}


\maketitle

\section{Introduction}
\label{sec_intro}

Pairwise similarity measurements of documents is a fundamental task in many text mining problems such as query-by-example, document classification and clustering. 

In the bag-of-words (BoW) \citep{IntroIR_Salton1986, IntroIR_Manning2008} vector space model, a document ${\bf x}$ is represented by an $M$-dimensional vector where $M$ is the number of terms in a given dictionary, i.e., ${\bf x}=\langle x_1, x_2, \cdots, x_M\rangle$; and it has the following two representations:
\begin{enumerate}
    \item Term-frequency-based representation: each $x_i\in \mathbb Z_+$ ($\mathbb Z_+$ is a set of non-negative integers) is the occurrence frequency of term $t_i$ in document ${\bf x}$.
    \item Binary representation: each $x_i\in \{0,1\}$ where 0 represents the absence of term $t_i$ in document ${\bf x}$ and 1 represents the presence of $t_i$ in ${\bf x}$.
\end{enumerate}

Because the number of terms in a document is significantly less than that in the dictionary, every document is represented as a sparse BoW vector, where many entries are zero. Because of sparsity, Euclidean distance is not a good similarity measure and the angular distance, aka cosine distance, is a preferred choice of inter-document similarity measure \citep{IntroIR_Salton1986, TFIDF_Salton1988}.

Because all terms in a document are not equally important to represent its subject, different `term weighting' schemes \citep{IntroIR_Manning2008, TFIDF_Salton1988} are used to adjust vector components based on the importance of their terms. 

The idea of term weighting was first introduced in the field of Information Retrieval (IR) where the task is to measure the relevance of documents in a given collection $D$ for a given query phrase consisting of a few terms. It is based on the following two assumptions \citep{IntroIR_Manning2008, TFIDF_Salton1988, SimilaritySpace_Zobel1998}: 
\begin{enumerate}
\item[i.] A term is important in a document if it occurs multiple times in the document.
\item[ii.] A rare term (that occurs in a few documents in the collection) is more important than frequent terms (that occur in many documents in the collection).
\end{enumerate}

The importance of terms in a document are estimated independent of the query. Because a query in the IR task is short and each term generally occurs only once, it is not an issue that the weights are determined independent of the query. 

However, it can be counter-productive in the query-by-example task where the query itself is a document, and terms often occur more than once in the query document. For example, to a query document ${\bf q}$, a document ${\bf x}$ having more occurrences of the terms in ${\bf q}$ may not be more similar than ${\bf y}$ which has exactly the same occurrences of terms in ${\bf q}$.
%It might be the case that ${\bf y}$ is exactly the same document as ${\bf x}$.

%The existing line of research in the tasks, which require inter-document similarity measurements, focuses
Prior research in the BoW inter-document similarity measurement task were focused on developing effective term weighting schemes to improve the task specific performances of existing measures such as cosine and Best Match 25 (BM25) \citep{TFIDF_Salton1988, Okapi_Robertson1994, ProbTFIDF_Joachims1997, PhDThesis_Singhal1997, ProbRelFramework_Robertson2009, IRWeightSenAna_Paltoglou2010, kNN_LSHTC_Han2012, ICF_Wang2013}. In contrast, we investigate an alternative similarity measure where an adjustment of vector components using term weighting is not required. 

This paper makes the following contributions:

\begin{enumerate}
\item Identify the shortcomings of the underlying assumptions of term weighting schemes employed in existing measures; and provide an alternative which is more congruous with the requirements of inter-document similarity measurements.
\item Introduce a new simple but effective inter-document similarity measure which is based on the new assumption and does not require explicit term weighting. It uses a more nuanced probabilistic approach than those used in term weighting to measure the similarity of two documents w.r.t each term occurring in the two documents under measurement.
\item Compare the performance of the new measure with existing measures (which use different term weighting schemes) in the query-by-example task. Our result reveals that the new measure produces (i) better results than existing measures in the binary BoW representation; and (ii) competitive and more consistent results to existing measures in the term-frequency-based BoW representation.
\end{enumerate}

The rest of the paper is organized as follows. Related work in the areas of term weighting and inter-document similarity measures are discussed in Section \ref{sec_relatedWork}. Issues of term weighting in inter-document similarity measurements are discussed Section~\ref{sec_Issues}. The proposed new inter-document similarity measure is presented in Section \ref{sec_sidf}, followed by empirical results in Section \ref{sec_exp}, related discussion in Section~\ref{sec_discussion}, and the last section presents the conclusions.

The key notations used in this paper are defined in Table~\ref{tbl_notations}.

\begin{table}
\caption{Key notations}
\centering
\begin{tabular}{l @{\hspace{20pt}} l}
\hline\hline 
$D$ & A collection of $N$ documents (i.e., $|D|=N$) \\
${\bf x}$ & BoW vector of a document $\langle x_1,x_2,\cdots,x_M\rangle$ \\
$t_i$ & The $i^{th}$ term in the dictionary \\
$n_i$ & The number of documents in $D$ having $t_i$ \\
%$x_i$ & The occurrence frequency of term $t_i$ in ${\bf x}$ \\
$T_{\bf x}$ & The set of terms in ${\bf x}$ \\
$w_i({\bf x})$ & The importance or weight of $t_i$ in ${\bf x}$ \\
$tf_i({\bf x})$ & Term frequency factor of $t_i$ in ${\bf x}$ \\
$idf(t_i)$ & Inverse document frequency factor of $t_i$ \\
$s({\bf x}, {\bf y})$ & The similarity of two documents ${\bf x}$ and ${\bf y}$ \\
%$P(e)$ & The Probability of an event $e$ happening \\
$dl({\bf x})$ & The length of document ${\bf x}$ (i.e., $\sum_{i=1}^M x_i$) \\
$avgdl$ & The average length of documents in $D$ \\
${\bf x} \onset{\succ}{{\bf q}\{i\}} {\bf y}$ & ${\bf x}$ is more similar to ${\bf q}$ than ${\bf y}$ w.r.t $t_i\in T_{\bf q}$ \\
${\bf x} \onset{=}{{\bf q}\{i\}} {\bf y}$ & ${\bf x}$ is equally similar to ${\bf q}$ as ${\bf y}$ w.r.t $t_i\in T_{\bf q}$ \\
\hline\hline
\end{tabular}
\label{tbl_notations}
\end{table}

%================ RELATED WORK ==============

\section{Related work}
\label{sec_relatedWork}

In this section, we present the pertinent details of term weighting and some widely used existing BoW inter-document similarity measures.

\subsection{{\bf Term weighting}}
\label{subsec_tf-idf}

In the field of IR, there has been considerable research investigating the effective term weighting scheme. The importance of a term $t_i$ in document ${\bf x}$, $w_i(\bf x)$, is estimated using different variants and combinations of two factors \citep{IntroIR_Manning2008, TFIDF_Salton1988, ProbTFIDF_Joachims1997, Okapi_Robertson1994, PhDThesis_Singhal1997, ProbRelFramework_Robertson2009, IRWeightSenAna_Paltoglou2010, kNN_LSHTC_Han2012, ICF_Wang2013}: (i) document-based factor based on the frequency of $t_i$ in ${\bf x}$, $x_i$; and (ii) collection-based factor based on the number of documents where $t_i$ occurs, $n_i$.

The most widely used term weighting scheme is tf-idf (term frequency - inverse document frequency) where $w_i({\bf x})=tf_i({\bf x}) \times idf(t_i)$ \citep{ IntroIR_Manning2008, TFIDF_Salton1988}; and it consists of:
\begin{enumerate}
    \item[i.] Document-based factor: $tf_i({\bf x})=1+\log(x_i)$ if $x_i>0$, and 0 otherwise;
    \item[ii.] Collection-based factor: $idf(t_i) = \log\left(\frac{N}{n_i}\right)$.
\end{enumerate}

In the IR task, the idea of tf-idf term weighting is based on the following assumptions \citep{SimilaritySpace_Zobel1998}:
\begin{enumerate}
    \item[i.] Documents with multiple occurrences of query terms are more relevant than documents with single occurrence of query terms [the tf assumption].
    \item[ii.] Documents having rare query terms (occurring  in a few documents in the collection) are more relevant to the query than documents having frequent query terms (occurring  in many documents in the collection) [the idf assumption].
\end{enumerate}

The tf factor considers the importance of $t_i$ in a document. Even though a document with multiple occurrences of a query term is more likely to be relevant to the given query, a document having higher occurrences of one query term is not necessarily more relevant than a document having lower occurrences of two query terms. Therefore, the logarithmic scaling of raw term frequencies is used to reduce the over influence of high frequencies of query terms \citep{IntroIR_Manning2008, TFIDF_Salton1988}. 

The idf factor considers the importance of $t_i$ in the given collection. Basically, it ranks the importance of terms in the given dictionary based on the number of documents where they occur. Terms occurring only in a few documents (i.e., rare terms) are considered to be more important in documents; and they are given more weights than the terms occurring in many documents (i.e., frequent terms) \citep{IntroIR_Manning2008, TFIDF_Salton1988}. 

\subsection{{\bf Inter-document similarity measures}}
\label{subsec_exisitingMeasures}

Here, we discuss three commonly used measures to estimate the similarity of two document vectors ${\bf x}$ and ${\bf y}$, $s({\bf x}, {\bf y})\to \mathbb R$  where $\mathbb R$ is a real domain.

\subsubsection{Cosine similarity}
\label{subsec_cosine}

The cosine similarity measure with the tf-idf term weighting is the most commonly used inter-document similarity measure. Using term weighted vectors, the cosine similarity of two documents ${\bf x}$ and ${\bf y}$ is estimated as:

\begin{equation}
\label{eqn_cos_sim}
s_{cos}({\bf x}, {\bf y}) = \frac{\sum_{i=1}^M w_i({\bf x})\times w_i({\bf y})}{\sqrt{\sum_{i=1}^M {w_i({\bf x})}^2}\times \sqrt{\sum_{i=1}^M {w_i({\bf y})}^2}}
\end{equation}

Note that the two terms in the denominator of Eqn~\ref{eqn_cos_sim} are the Euclidean lengths ($\ell_2$-norms) of the term weighted vectors. 

It is important to normalize the similarity of documents by their lengths, otherwise it favors longer documents which have higher probability of having more terms in common with the query document over shorter documents \citep{IntroIR_Salton1986, IntroIR_Manning2008, TFIDF_Salton1988, PivotedLenghtNormalisation_Singhal1996}.

\subsubsection{Best Match 25 (BM25)}
\label{subsec_bm25}

BM25 \citep{ProbRelFramework_Robertson2009, ProbIR_Jones2000} is the state-of-the-art document ranking measure in IR. It is based on the probabilistic framework of term weighting by \cite{Okapi_Robertson1994}. \cite{kNN_LSHTC_Han2012} used BM25 to measure the similarity of two documents ${\bf x}$ and ${\bf y}$ as follows:

\begin{equation}
    \begin{multlined}
    s_{bm25}({\bf x},{\bf y}) = \displaystyle \sum_{i=1}^M idf_{bm25}(t_i) \times \frac{x_i\cdot (a + 1)}{x_i + a\cdot\left(1-b+b\cdot\frac{dl({\bf x})}{avgdl}\right)} \times \\ \frac{y_i\cdot (a + 1)}{y_i + a\cdot\left(1-b+b\cdot\frac{dl({\bf y})}{avgdl}\right)}
    \end{multlined}
    \label{eqn_bm25}
\end{equation}

\noindent where $dl({\bf x})=\sum_{i=1}^M x_i$ is the normal length of document ${\bf x}$ (i.e., $\ell_1$-norm of the unweighted vector); $avgdl=\frac{1}{N}\sum_{{\bf x}\in D}dl({\bf x})$ is the average normal document length; $a$ and $b$ are free parameters that control the influence of the term frequencies and document lengths; and $idf_{bm25}(t_i)$ is the idf factor of term $t_i$ defined as follows:
\begin{equation}
    idf_{bm25}(t_i) = \log \left(\frac{N-n_i+0.5}{n_i+0.5}\right)
    \label{eqn_idf_bm25}
\end{equation}

It uses different variants of tf and idf factors in the similarity measure. The pivoted normal document length \citep{PivotedLenghtNormalisation_Singhal1996} is used in the tf factor so that longer documents which have higher probability of having more terms in common with the query document are not favored over shorter documents.

\subsubsection{Jaccard similarity}
\label{subsec_jaccard}

The Jaccard similarity \citep{Jaccard1901} of two documents ${\bf x}$ and ${\bf y}$ is estimated as follows:
 
 \begin{equation}
     s_{jac}({\bf x}, {\bf y}) = \frac{|T_{\bf x} \cap T_{\bf y}|}{|T_{\bf x} \cup T_{\bf y}|}
     \label{eqn_jaccardSim}
 \end{equation}

\noindent where $T_{\bf x} = \{t_i: x_i > 0\}$ is the set of terms in document ${\bf x}$ and $|\cdot|$ is the cardinality of a set. 

It only considers the number of terms occurring in both ${\bf x}$ and ${\bf y}$ and does not take into account the importance of terms in documents. The similarity is normalized by the number of distinct terms occurring in either ${\bf x}$ or ${\bf y}$ to take into account that ${\bf x}$ and ${\bf y}$ have higher chance of having terms in common if they have more terms. 

The weighted or generalized version of Jaccard similarity \citep{WeightedJacard_Chierichetti2010} of two documents using term weighted vectors is defined as follows:

 \begin{equation}
     s_{wjac}({\bf x}, {\bf y}) = \frac{\sum_{i=1}^M\min\{w_i({\bf x}), w_i({\bf y})\}}{\sum_{i=1}^M\max\{w_i({\bf x}), w_i({\bf y})\}}
  \label{eqn_wtJaccardSim}
 \end{equation}
 
The similarity of ${\bf x}$ and ${\bf y}$ w.r.t $t_i\in T_{\bf x} \cap T_{\bf y}$ depends on the importance of $t_i$ in the two documents. The similarity is normalized by the sum of maximum weights of all $t_i\in T_{\bf x} \cup T_{\bf y}$.

Note that the weighted Jaccard similarity of ${\bf x}$ and ${\bf y}$ (Eqn~\ref{eqn_wtJaccardSim}) in the binary BoW vector representation without any term weighting is equivalent to the traditional Jaccard similarity (Eqn~\ref{eqn_jaccardSim}).

%============= Issues of tf-idf in inter-document similarity measurements ==============

\section{Issues of the tf-idf assumptions in inter-document similarity measurements}
\label{sec_Issues}

Even though the tf and idf assumptions discussed in Section~\ref{subsec_tf-idf} are intuitive in the IR  task to rank documents for a given query phrase of a few terms, they can be counter-intuitive in the query-by-example task which requires inter-document similarity measurements to rank documents in $D$ w.r.t a given query document. 

In the literature, the query-by-example task is treated as the IR task where query is a document; and the same idea of the tf-idf term weighting is used. However, there is a fundamental difference between the two tasks --- unlike in the typical IR task where the query comprises of a few distinct terms (i.e., each term generally occurs only once in the query phrase), the query in the query-by-example task is a long document which often has multiple occurrences of terms.

\subsection{{\bf Issue of the tf assumption}}
\label{subsec_tf-issue}

For a query document ${\bf q}$ with terms $T_{\bf q}$, a document ${\bf x}$ having more occurrences of terms in $T_{\bf q}$ than in ${\bf q}$, may not be more similar to ${\bf q}$ than another document ${\bf y}$, which has similar occurrences of terms in $T_{\bf q}$ as in ${\bf q}$. For example, let's assume ${\bf x}$ and ${\bf y}$ have frequencies of $t_r\in T_{\bf q}$ as $x_r=10$ and $y_r=1$, respectively. If ${\bf q}$ has $q_r=1$, it is difficult to say that ${\bf x}$ is more similar to ${\bf q}$ than ${\bf y}$ w.r.t $t_r$ $\left(i.e., {\bf x} \onset{\succ}{{\bf q}\{r\}} {\bf y}\right)$, simply because of $x_r>q_r$ (and $q_r=y_r$). It might be the case that ${\bf y}$ is exactly the same document as ${\bf q}$. 

Because of the tf-based term weighting factor, ${\bf x}\neq {\bf q}$ can be more similar to ${\bf q}$ than ${\bf q}$ itself using some existing measure such as BM25\footnote{It depends on the lengths of documents and parameters $a$ and $b$.}. Thus, the tf assumption can be counter-intuitive in inter-document similarity measurements. 

\subsection{{\bf Issue of the idf assumption}}
\label{subsec_idf-issue}
Similarly, ${\bf x}$ having rare terms of $T_{\bf q}$ may not be more similar to ${\bf q}$ than ${\bf y}$ having frequent terms of $T_{\bf q}$. For example, let's assume the scenario presented in Table~\ref{tbl_scenario}: 

\begin{table}[!htb]
\centering
\caption{A scenario to demonstrate the issue of the idf assumption. Note that all $\frac{N}{2}$ documents having $t_g$ have its frequency of 1; and all $N$ documents having $t_h$ have its frequency of 1 except ${\bf y}$ where $y_h=10$}
\begin{tabular}{r @{\hspace{25pt}} r @{\hspace{20pt}} r @{\hspace{20pt}} r @{\hspace{20pt}} r @{\hspace{25pt}} r}
\hline\hline
 & $n$ & $idf(t)$ & $\bf x$ & $\bf y$ & $\bf q$ \\ 
 \hline\hline
 {$t_g$} & $\frac{N}{2}$ & $\log(2)$ & 1 & 1 & 1\\
 {$t_h$} & $N$ & 0 & 1 & 10 & 10 \\
\hline\hline
\end{tabular}
\label{tbl_scenario}
\end{table}

\begin{table}[t]
\centering\small
\caption{The tf-idf weighting (in existing measures) versus Sp: (i) Underlying assumptions for documents to be relevant/similar to a query document ${\bf q}$; and the relation of similarities of ${\bf x}$ and ${\bf y}$ to ${\bf q}$ (ii) in the same example discussed in Section~\ref{subsec_tf-issue} and (iii) in the same example used in Section \ref{subsec_idf-issue}.}
\begin{tabular}{@{} r @{\hspace{3pt}} l @{\hspace{2pt}} l l @{\hspace{3pt}} l @{}}
\hline\hline
\multicolumn{3}{l}{tf-idf term weighting} & \multicolumn{2}{l}{Sp (the proposed approach)} \\ 
\hline\hline
\multicolumn{5}{@{\hspace{1pt}}l}{(i) Underlying Assumptions} \\ \hline
tf: & ${\bf y} \onset{\succ}{{\bf {q}}\{i\}} {\bf x}$ & if $y_i > x_i$ & ${\bf y} \onset{\succ}{{\bf q}\{i\}} {\bf x}$ & if $|\{{\bf z} \in D:\alpha(y_i,q_i) \le z_i \le \beta(y_i,q_i)\}|<$ \\
idf: & ${\bf y} \onset{\succ}{{\bf {q}}\{i,j\}} {\bf x}$ & if $n_i < n_j$ (for $q_i>0, q_j>0$;& & \hspace{7pt} $|\{{\bf w} \in D:\alpha(x_i,q_i) \le w_i \le \beta(x_i,q_i)\}|$ \\  
\multicolumn{3}{l}{\hspace{30pt} $x_i=0, x_j=q_j$; $y_i=q_i, y_j=0$)} & & \hspace{-7pt}where $\alpha(\cdot,\cdot)=\min(\cdot,\cdot); \beta(\cdot,\cdot)=\max(\cdot,\cdot)$\\ 
\hline
\multicolumn{5}{@{\hspace{1pt}}l}{(ii) Example discussed in Section~\ref{subsec_tf-issue} ($x_r=10, y_r=1, q_r=1$)} \\ \hline
\multicolumn{3}{l}{${\bf x} \onset{\succ}{{\bf q}\{r\}} {\bf y}$ \hspace{13pt} because $x_r>y_r$ (even though} & ${\bf y} \onset{\succ}{{\bf q}\{r\}} {\bf x}$ & because $|\{{\bf z} \in D:q_r = z_r = y_r\}|<$\\
& & \hspace{80pt} $y_r=q_r$) & & \hspace{32pt} $|\{{\bf w} \in D:q_r \le w_r \le x_r\}|$\\ \hline
\multicolumn{5}{@{\hspace{1pt}}l}{(iii) Example discussed in Section~\ref{subsec_idf-issue} (Table~\ref{tbl_scenario})} \\ \hline
\multicolumn{3}{l}{${\bf x} \onset{=}{{\bf q}\{g,h\}} {\bf y}$ \hspace{6pt} because (i) $q_g=x_g=y_g=1$;} & ${\bf y} \onset{\succ}{{\bf q}\{g,h\}} {\bf x}$ & because (i) $|\{{\bf z} \in D:q_g =  z_g = y_g\}|=$\\
& &  \hspace{50pt} and & & \hspace{45pt} $|\{{\bf w} \in D:q_g = w_g = x_g\}|$\\
& & (ii) $idf(t_h)=0$ (even though & & \hspace{30pt} (ii) $|\{{\bf z} \in D:q_h =  z_h = y_h\}|<$\\
& & \hspace{7pt} $q_h=y_h=10$ and $x_h=1$) & & \hspace{45pt} $|\{{\bf w} \in D:x_h \le w_h \le q_h\}|$\\
\hline\hline
\end{tabular}
\label{tbl_compare}
\end{table}

Because of $idf(t_h)=0$, the term $t_h$ will be completely ignored. However, $q_h=y_h=10$ is more useful than $q_g=x_g=y_g=1$ because ${\bf y}$ is the only document in $D$ which has as many occurrences of $t_h$ as ${\bf q}$. Even though there is no discrimination between documents w.r.t $t_g$ (all $\frac{N}{2}$ documents with $t_i$ have its frequency of 1), $t_g$ is assigned more weight with $idf(t_g)=\log 2$ than $t_h$ with $idf(t_h)=0$. As a result, ${\bf x}$ and ${\bf y}$ become equally similar to ${\bf q}$ w.r.t $t_g$ and $t_h$ $\left(i.e., {\bf x} \onset{=}{{\bf q}\{g,h\}} {\bf y}\right)$ even though ${\bf y}$ has exactly the same occurrences of $t_g$ and $t_h$ as ${\bf q}$. This example shows that the idf assumption can be counter-intuitive in document similarity measurements.

%============= Our New Proposal (Sp) ==============

\section{Our proposal to overcome the issues of tf-idf based term weighting in document similarity measurements}
\label{sec_sidf}

The main problem of the tf-idf term weighting in inter-document similarity measurements is that the importance of $t_i$ in ${\bf x}$, $w_i({\bf x})$, is estimated without considering the frequency of $t_i$ in ${\bf q}$, $q_i$. It is not an issue in the IR task because $q_i$ is almost always 1 if $t_i$ occurs in the given query phrase ${\bf q}$. In a query document, $q_i$ can be larger than 1. Therefore, judging the importance of $t_i$ in ${\bf x}$, without considering $q_i$, can be counter-productive in inter-document similarity measurements.

A more fit-to-the-purpose approach would be to evaluate the importance of $t_i$ in ${\bf x}$ by examining the similarity of $x_i$ w.r.t. $q_i$. However, as discussed in Section~\ref{subsec_idf-issue}, simply having similar occurrences of $t_i$ (i.e., $x_i=q_i$) is not sufficient to consider them to be similar. The similarity shall also consider how rare the frequency of $t_i$ is in the collection. 

Putting the above requirements together, the similarity of ${\bf x}$ and ${\bf q}$ w.r.t $t_i$ shall be based on the number of documents in $D$ which have similar occurrence frequencies of $t_i$ as in both ${\bf x}$ and ${\bf q}$. More formally, ${\bf x}$ and ${\bf q}$ are more likely to be similar w.r.t $t_i$ if $|\{{\bf z} \in D: \min(x_i, q_i) \leq z_i \leq \max(x_i, q_i) \}|$ is small. The first part in Table~\ref{tbl_compare} compares the underlying assumptions of the tf-idf term weighting (used in existing measures) and the proposed approach called Sp, to be introduced in the next subsection.

This approach addresses the limitations of both the tf and idf assumptions discussed in Section~\ref{sec_Issues}. The results of the new approach using the same examples discussed in Sections~\ref{subsec_tf-issue} and \ref{subsec_idf-issue} are provided in the second part of Table~\ref{tbl_compare}. The comparisons demonstrate that the new approach provides more intuitive outcomes than the tf-idf term weighting.

%In the example discussed in Section~\ref{subsec_tf-issue}, ${\bf y} \onset{\succ}{{\bf q}\{i\}} {\bf x}$ because $|\{{\bf z} \in D:1 \leq z_i\leq 1\}| < |\{{\bf w} \in D: 1 \leq w_i \leq 10 \}|$. Similarly, in the example discussed in Section~\ref{subsec_idf-issue}, because $|\{{\bf z} \in D:10 \leq z_j \leq 10 \}| < |\{{\bf w} \in D: 1 \leq w_i\leq 1\}|$, $q_j=x_j=10$ has more influence than $q_i=x_i=1$ in the similarity of ${\bf x}$ and ${\bf q}$.

\subsection{{\bf Sp: A new document similarity measure}}

Recently, \cite{mp_Aryal2014, MpKAIS_Aryal2017} introduced a data dependent measure where the similarity of two data objects $u$ and $v$ depends on the distribution of data between $u$ and $v$ \citep{mp_Aryal2014, MpKAIS_Aryal2017}. The intuition is that $u$ and $v$ are more likely to be similar if there are less data between them. For example, two individuals earning 800k and 900k are judged to be more similar by human than two individuals earning 50k and 150k because a lot more people earn in [50k, 150k] than [800k, 900k]. 

Using the similar idea, the similarity of two documents ${\bf x}$ and ${\bf y}$ can be estimated as: 
\begin{equation}
    %s_{sp}({\bf x}, {\bf y}) = \frac{1}{|T_{\bf x} \cup T_{\bf y}|} \sum_{t_i \in T_{\bf x} \cap T_{\bf y}} \log\frac{N}{|\{{\bf z} \in D:\alpha_i\leq z_i\leq \beta_i \}|}
    s_{sp}({\bf x}, {\bf y}) = \frac{1}{|T_{\bf x} \cup T_{\bf y}|} \sum_{t_i \in T_{\bf x} \cap T_{\bf y}} \log\frac{N}{|\{{\bf z} \in D:\min(x_i, y_i)\leq z_i\leq \max(x_i, y_i) \}|}
    \label{eqn_sidf}
\end{equation}

\noindent where 
%$\alpha_i = \min(x_i, y_i)$ and $\beta_i = \max(x_i, y_i)$; and 
$\frac{1}{|T_{\bf x} \cup T_{\bf y}|}$ is a normalization term to account for the probability of a term occurring in both ${\bf x}$ and ${\bf y}$. It reduces the bias towards documents having more terms because they have a higher probability of having terms, which also exist in a query document, than documents having fewer terms. 

The number of distinct terms is used as a normalization factor (as in the traditional Jaccard similarity) because it is not sensitive to multiple occurrences of the terms in a document which do not occur in the query document. In the IR task, \cite{PivotedLenghtNormalisation_Singhal1996} have shown that it is more effective than the cosine or normal length normalizations which penalize more to documents having multiple occurrences of the terms which are not in the query phrase.

Sp can be interpreted as a simple probabilistic measure where the similarity of two documents w.r.t $t_i \in T_{\bf x} \cap T_{\bf y}$ is assigned based on the probability of the frequency of $t_i$ to be in [$\min(x_i, y_i)$, $\max(x_i, y_i)$], $P(\min(x_i, y_i) \leq \chi_i \leq \max(x_i, y_i))$ (where $\chi_i$ be a random variable representing the occurrence frequency of term $t_i$ in a document). In practice, $P(\min(x_i, y_i) \leq \chi_i \leq \max(x_i, y_i)) = \frac{|\{{\bf z} \in D\mbox{ }:\mbox{ }\min(x_i, y_i)\leq z_i\leq \max(x_i, y_i) \}|}{N}$, which is the inverse of the term used in Eqn~\ref{eqn_sidf}. 

\subsection{{\bf Characteristics of Sp}}

The proposed measure has the following characteristics:

\renewcommand{\labelenumi}{\roman{enumi})}

\begin{enumerate}

\item {\it Term weighting is not required}:

Unlike in existing measures such as cosine and BM25, $x_i$ and $y_i$ are not used directly in the similarity measure. They are used just to define $\min(x_i, y_i)$ and $\max(x_i, y_i)$; and the similarity is based on $|\{{\bf z} \in D:\min(x_i, y_i)\leq z_i\leq \max(x_i, y_i)\}|$ which is invariant to the monotonic scaling of frequency values. Hence, Sp does not require additional term weighting to adjust frequency values. \\

\item {\it Self-similarity is data dependent and the upper bound of similarity}:

Unlike cosine and both variants of Jaccard similarity where the self-similarity of documents is fixed with the maximum of 1, Sp has data dependent self similarity because $s_{sp}({\bf x}, {\bf x})$ depends on the $P(x_i)$ for all $t_i \in T_{\bf x}$. Thus, $s_{sp}({\bf x}, {\bf x})$ and $s_{sp}({\bf y}, {\bf y})$ can be different.

The similarity in Sp is bounded by its self-similarity i.e., $\forall_{{\bf y} \neq {\bf x}} \mbox{ } s_{sp}({\bf x}, {\bf x}) > s_{sp}({\bf x}, {\bf y})$. Although BM25 also has data dependent self-similarity, it is possible to have similarity of different documents to be larger than the self-similarity, i.e., there may be ${\bf y}\neq {\bf x}$ with $s_{bm25}({\bf x}, {\bf y}) > s_{bm25}({\bf x}, {\bf x})$\footnote{It depends on the lengths of ${\bf x}$ and ${\bf y}$ and parameters $a$ and $b$.}.\\

\item {\it Relationship with the traditional Jaccard similarity and idf term weighting}:

The formulation of Sp (Eqn~\ref{eqn_sidf}) looks similar to the formulation of the traditional Jaccard similarity (Eqn~\ref{eqn_jaccardSim}) except that the similarity of ${\bf x}$ and ${\bf y}$  w.r.t $t_i \in T_{\bf x} \cap T_{\bf y}$ is based on $|\{{\bf z} \in D:\min(x_i,y_i)\leq z_i\leq \max(x_i,y_i)\}|$ in Sp, whereas it is the fixed constant of 1 in the traditional Jaccard similarity. 

In the binary BoW vector representation, when $t_i \in T_{\bf x} \cap T_{\bf y}$ and $|\{{\bf z} \in D: z_i = 1\}|=n_i$, Sp assigns the similarity of ${\bf x}$ and ${\bf y}$  w.r.t $t_i$ based on $idf(t_i)$, whereas in the traditional Jaccard similarity, it is 1, irrespective of whether $t_i$ is rare or frequent in $D$. 

In the term-frequency-based BoW representation, Sp is different from the idf weighting because $idf(t_i)$ is based on $|\{{\bf z} \in D:z_i > 0\}|$, whereas Sp is based on $|\{{\bf z} \in D:\min(x_i,y_i) \leq z_i\leq \max(x_i, y_i)\}|$, where $x_i>0$ and $y_i > 0$ when $t_i \in T_{\bf x} \cap T_{\bf y}$. 

\end{enumerate}
 
\subsection{{\bf Computational complexity}}
\label{subsec_sidf_complexity}

In the term-frequency-based BoW vector representation, it appears that computing $|\{{\bf z} \in D:\min(x_i,y_i) \leq z_i\leq \max(x_i, y_i)\}|$ naively can be expensive as it requires a range search to find the number of documents having the frequencies of $t_i$ between $x_i$ and $y_i$. Since, all $x_i$ are integers (term occurrence frequency counts), it can be computed in constant time by the following simple preprocessing. 

Let $m_i$ be the maximum occurrence frequency of term $t_i$ in any document in the given collection $D$. We can maintain a cumulative frequency count array $F_i$ of size $m_i+1$ where $F_i[j]$ contains the number of documents having occurrences of $t_i$ fewer than or equal to $j$. 

Using $F_i$, it can be estimated in constant time as $|\{{\bf z} \in D:\min(x_i,y_i) \leq z_i\leq \max(x_i, y_i)\}|= F_i[\max(x_i, y_i)]-F_i[\min(x_i, y_i)-1]$. Note that $\min(x_i, y_i)$ can not be 0 because $|\{{\bf z} \in D:\min(x_i,y_i) \leq z_i\leq \max(x_i, y_i)\}|$ is computed only if $t_i \in T_{\bf x} \cap T_{\bf y}$ (i.e., $x_i>0$ and $y_i>0$) and thus $\min(x_i, y_i) > 0$.

The above preprocessing requires $O(MN)$ time complexity and $O(Mm)$ space complexity, where $m$ is the average maximum frequency of terms. 

With the above preprocessing, the runtime of computing the similarity of ${\bf x}$ and ${\bf y}$ using Sp is the same as that of the existing similarity measures which is $O(M)$.

%==== EMPIRICAL RESULTS =====

\section{Empirical evaluation}
\label{sec_exp}

In this section, we present the results of experiments conducted to evaluate the task specific performances of Sp, BM25, weighted Jaccard and cosine similarity in the query-by-example task to retrieve documents similar to a given query document. We did experiments with both the term-frequency-based and binary BoW vector representations. We used different combinations of tf and idf based term weighting factors with the weighted Jaccard and cosine similarity measures.

\subsection{{\bf Datasets and experimental setup}}

We used 10 datasets from 6 benchmark document collections. The characteristics of data sets are provided in Table~\ref{tbl_data}. NG20, R8, R52 and Webkb are from \cite{PhDThesis_Ana2007}\footnote{BoW vectors available at: http://web.ist.utl.pt/acardoso/datasets/}; and others are from \cite{Centroid_Han2000}\footnote{BoW vectors available at: http://www.cs.waikato.ac.nz/ml/weka/datasets.html}.

\begin{table}[t]
\centering
\caption{Characteristics of datasets ($N$: Number of documents, $M$: Number of terms, $C$: Number of classes).}
\begin{tabular}{ l @{\hspace{25pt}} r @{\hspace{25pt}} r @{\hspace{25pt}} r @{\hspace{25pt}} l }
\hline\hline
Name    &    $N$  &     $M$    &    $C$  &   Collection \\ 
\hline\hline
Fbis   &   2,463  &   2,000    &	17   &   TREC collection \\
La1s   &   3,204  &  13,195    &     6   &   TREC collection \\
La2s   &   3,075  &  12,432    &     6   &   TREC collection \\
New3s  &   9,558  &  26,832    &    44   &   TREC collection \\
Ng20   &  18,821  &   5,489    &    20   &   20 Newsgroup collection \\
Ohscal &  11,162  &  11,465    &    10   &   Ohsumed patients records \\
R8     &   7,674  &   3,497    &     8   &   Reuters collection \\
R52    &   9,100  &   7,379    &    52   &   Reuters collection \\
Wap    &   1,560  &   8,460    &    20   &   Yahoo web pages \\
Webkb  &   4,199  &   1,817    &     4   &   University web pages \\
\hline\hline
\end{tabular}
\label{tbl_data}
\end{table}

Each dataset was divided into two subsets ${\mathcal D}$ and ${\mathcal Q}$ using a 10-fold cross validation which have 90\% and 10\% of the documents, respectively. ${\mathcal D}$ was used as a given collection from which similar documents were extracted for each query document in ${\mathcal Q}$. For each ${\bf q}\in {\mathcal Q}$, documents in ${\mathcal D}$ were ranked in descending order of their similarities to ${\bf q}$ using different contending similarity measures. The top $k$ documents were presented as the similar documents to  ${\bf q}$. 

For performance evaluation, a document was considered to be similar to ${\bf q}$ if they have the same class label. In order to demonstrate the consistency of a measure at different top $k$ retrieved results, we evaluated the precision at the top $k$ retrieved results ($P@k$ in terms of percentage) with  $k=1,2,\cdots,25$ and used the mean average precision up to $k=25$. The performance evaluation criterion is: $MAP@25=\frac{\sum_{k=1}^{25} P@k}{25}$. 

We repeated the experiment 10 times using each of the 10 folds as ${\mathcal Q}$ and the remaining 9 folds as ${\mathcal D}$. The average $MAP@25$ and standard error over the 10 runs were reported. The average $MAP@25$ of two measures were considered to be significantly different if their confidence intervals based on two standard errors were not overlapping.

The free parameters $a$ and $b$ in BM25 were set to 1.2 and 0.95, respectively, as recommended by \cite{IRWeightSenAna_Paltoglou2010} and \cite{ProbIR_Jones2000}.

All the experimental setups and similarity measures were implemented in Java using the WEKA platform \citep{Weka_2009}. All the experiments were conducted in a Linux machine with 2.27 GHz processor and 16 GB memory. We discuss the experimental results with the term-frequency-based and binary BoW vector representations separately in the following two subsections.

\subsection{{\bf Results in the term-frequency-based BoW vector representation}}

Here we used two term weighting schemes - tf factor only and tf-idf factors, with weighted Jaccard and cosine. The six contending measures are: {\it Sp}, {\it BM25}, {\it Cos.tf-idf} (cosine with tf-idf), {\it Cos.tf} (cosine with tf only), {\it WJac.tf-idf} (weighted Jaccard with tf-idf) and {\it WJac.tf} (weighted Jaccard with tf only). 

The average $MAP@25$ and standard error over 10 runs of six contending measures are provided in Table~\ref{tbl_tf_mapl} and the summarized results in terms of pairwise win-loss-draw counts of contending measures based on the two standard errors significance test over the 10 datasets used in the experiment are provided in Table~\ref{tbl_tf_wld}.

\begin{table}[t]
\centering\small
\caption{Term-frequency-based BoW representation: Average $MAP@25$ and standard error over 10 runs. The best result is underlined and the results equivalent (insignificant difference based on two standard errors) to the best result are bold faced.}
\begin{tabular}{l c c c c c c }
  \hline\hline
  & {\it BM25}    & {\it Cos.tf-idf}     & {\it Cos.tf}    & {\it WJac.tf-idf}    & {\it WJac.tf}   & {\it Sp} \\
  \hline\hline  
  Fbis    &  65.12$\pm$0.62                 &  {\bf 68.42$\pm$0.61}  &  {\bf 68.28$\pm$0.58}            &  \underline{\bf 68.48$\pm$0.49}  &  66.75$\pm$0.54                &  {\bf 67.77$\pm$0.51} \\ 
  La1s    &  74.41$\pm$0.32                 &  75.97$\pm$0.42        &  73.08$\pm$0.49                  &  {\bf 79.18$\pm$0.33}            &  77.54$\pm$0.47                &  \underline{\bf 79.36$\pm$0.32} \\  
  La2s    &  76.42$\pm$0.49                 &  78.11$\pm$0.42        &  75.24$\pm$0.44                  &  \underline{\bf 81.06$\pm$0.42}  &  79.45$\pm$0.37                &  {\bf 80.89$\pm$0.40} \\ 
  New3s   &  67.01$\pm$0.18                 &  68.31$\pm$0.19        &  \underline{\bf 70.19$\pm$0.19}  &  69.36$\pm$0.16                  &  68.45$\pm$0.15                &  68.98$\pm$0.16 \\  
  Ng20    &  \underline{\bf 76.47$\pm$0.19} &  74.81$\pm$0.24        &  67.80$\pm$0.28                  &  73.67$\pm$0.23                  &  64.28$\pm$0.24                &  72.30$\pm$0.20 \\ 
  Ohscal  &  59.72$\pm$0.22                 &  53.59$\pm$0.21        &  \underline{\bf 61.06$\pm$0.26}  &  59.68$\pm$0.21                  &  60.81$\pm$0.20                &  60.14$\pm$0.19 \\ 
  R52     &  85.50$\pm$0.20                 &  80.80$\pm$0.27        &  \underline{\bf 86.57$\pm$0.15}  &  84.55$\pm$0.21                  &  84.72$\pm$0.19                &  84.39$\pm$0.22 \\ 
  R8      &  91.05$\pm$0.14                 &  86.14$\pm$0.22        &  \underline{\bf 92.93$\pm$0.19}  &  91.03$\pm$0.18                  &  91.94$\pm$0.21                &  91.40$\pm$0.17 \\ 
  Wap     &  19.67$\pm$0.42                 &  65.33$\pm$0.34        &  61.97$\pm$0.41                  &  {\bf 70.54$\pm$0.46}            &  65.10$\pm$0.48                &  \underline{\bf 70.92$\pm$0.50} \\  
  Webkb   &  70.28$\pm$0.23                 &  68.55$\pm$0.24        &  73.04$\pm$0.27                  &  73.90$\pm$0.31                  &  \underline{\bf 75.25$\pm$0.25}&  {\bf 74.91$\pm$0.33} \\
\hline\hline
\end{tabular}
\label{tbl_tf_mapl}
\end{table}

\begin{table}[t]
\centering\small
\caption{Term-frequency-based BoW representation: Win-loss-draw counts of measures in columns against those in rows based on the two standard error significance test over 10 runs.}
%\begin{tabular}{@{\hspace{2pt}} l @{\hspace{4pt}} | c @{\hspace{8pt}} c @{\hspace{8pt}} c @{\hspace{8pt}} c @{\hspace{8pt}} c @{\hspace{2pt}}}
\begin{tabular}{l c c c c c }
\hline\hline
        &  {\it Sp} &  {\it WJac.tf}  &  {\it WJac.tf-idf}   & {\it Cos.idf}  &   {\it Cos.tf-idf}  \\
\hline\hline	       
{\it BM25}  &  8-2-0  &  8-2-0  &  6-2-2  &  7-0-3  &  5-5-0 \\
{\it Cos.tf-idf}   &  8-1-1  &  6-2-2  &  8-1-1  &  5-4-1 \\ 	
{\it Cos.tf}  &  5-4-1  &  4-5-1  &  5-4-1 \\		
{\it WJac.tf-idf}  &  3-2-5  &  3-6-1 \\			
{\it WJac.tf} &  5-2-3 \\				
\hline\hline
\end{tabular}
\label{tbl_tf_wld}
\end{table}

Table~\ref{tbl_tf_mapl} shows that {\it Sp} and {\it Cos.tf} produced the best or competitive to the best result in five datasets each; followed by {\it WJac.tf-idf} in four; whereas {\it Cos.tf-idf}, {\it BM25} and {\it WJac.tf} were best or competitive to the best measure in only one dataset each.

The first column in Table~\ref{tbl_tf_wld} shows that {\it Sp} had more wins than losses over all contending measures. It had one more wins than losses against the closest contenders {\it Cos.tf} and {\it WJac.tf-idf}.

Of the two cosine measures, {\it Cos.tf} had more wins than losses to {\it Cos.tf-idf}. This shows that the idf term weighting can be counter-productive with cosine in inter-document similarity measurements. It is mainly due to the cosine normalization which penalizes more to documents having rare terms (with high idf weights) which are not in ${\bf q}$. In comparison to {\it BM25}, {\it Cos.tf} produced better results with seven wins and no loss; and  {\it Cos.tf-idf} was competitive with five wins versus five losses. 

It is interesting to note that, in the Wap dataset, {\it BM25} produced significantly worse result than other contenders. It is due to the idf factor used in {\it BM25}. If a term $t_i$ occurs in more than half of the documents in ${\mathcal D}$ (i.e., $n_i>\frac{N}{2}$), $idf_{bm25}(t_i)$ is negative and $t_i$ has negative contribution in the similarity of two documents. When $idf_{bm25}(t_i)$ was replaced by the traditional $idf(t_i)$ in the formulation of {\it BM25} (Eqn~\ref{eqn_bm25}), it produced $MAP@25$ = 67.04\% which was still worse than those of {\it Sp} and {\it WJac.tf-idf}.  

In weighted Jaccard similarity, {\it WJac.tf-idf}  produced better retrieval results than {\it WJac.tf}. It is interesting to note that {\it WJac.tf-idf} produced better retrieval results than {\it Cos.tf-idf}, {\it Cos.tf} and {\it BM25}. This could be mainly due to the vector length normalization used in {\it BM25} and {\it cosine} that penalizes more to documents having higher frequencies of terms which are not in ${\bf q}$. 

It is interesting to note that {\it Sp} and {\it WJac.tf-idf} produced more consistent results than the other contending measures. They did not produce the worst result in any dataset whereas {\it WJac.tf} produced the worst result in one dataset (NG20) followed by {\it Cos.tf} in two datasets (La1s and La2s); {\it BM25} in three datasets (Fbis, New3s and Wap); and {\it Cos.tf-idf} in four datasets (Ohscal, R8, R52 and Webkb).

In terms of runtime, all measures had runtime in the same order of magnitude. For example, in the NG20 dataset, the average total runtime of one run (including preprocessing) using {\it Sp} took 15935 seconds; whereas {\it BM25}, {\it Cos.tf-idf} and {\it WJac.tf-idf} took 27432, 16089 and 14875 seconds, respectively.  

\subsection{{\bf Results in the binary BoW vector representation}}

Here, six contending measures are: {\it Sp}, {\it BM25}, {\it Cos.idf} (cosine with idf), {\it Cos} (cosine without idf), {\it WJac.idf} (weighted Jaccard with idf) and {\it WJac} (weighted Jaccard without idf). Note that {\it WJac} which is not using any term weighting is equivalent to the traditional Jaccard similarity defined in Eqn~\ref{eqn_jaccardSim}. 

The average $MAP@25$ and standard error over 10 runs of the six contending measures are provided in Table~\ref{tbl_bin_mapl}; and the summarized results in terms of pairwise win-loss-draw counts of contending measures based on the two standard errors significance test over the 10 datasets used in the experiment are provided in Table~\ref{tbl_bin_wld}.

\begin{table*}[t]
\centering\small
\caption{Binary BoW representation: Average $MAP@25$ and standard error over 10 runs. The best result is underlined and the results equivalent (insignificant difference based on two standard errors) to the best result are bold faced.}
\begin{tabular}{l c c c c c c }
\hline\hline
  & {\it BM25}  & {\it Cos.idf}   & {\it Cos}   & {\it WJac.idf}   & {\it WJac}  & {\it Sp} \\ 
  \hline\hline
  Fbis    &  \underline{\bf 67.90$\pm$0.50}  &  66.46$\pm$0.50  &  63.24$\pm$0.56                  &  {\bf 67.17$\pm$0.46}            &  64.58$\pm$0.52       &  {\bf 66.94$\pm$0.47} \\  
  La1s    &  74.78$\pm$0.25                  &  76.78$\pm$0.34  &  75.96$\pm$0.38                  &  {\bf 78.54$\pm$0.34}            &  77.55$\pm$0.39       &  \underline{\bf 79.04$\pm$0.30} \\  
  La2s    &  76.71$\pm$0.48                  &  78.48$\pm$0.42  &  77.55$\pm$0.38                  &  {\bf 80.02$\pm$0.39}            &  79.12$\pm$0.35       &  \underline{\bf 80.54$\pm$0.40} \\ 
  New3s   &  \underline{\bf 69.61$\pm$0.20}  &  66.73$\pm$0.16  &  64.88$\pm$0.15                  &  67.76$\pm$0.15                  &  65.66$\pm$0.16       &  68.13$\pm$0.16 \\  
  Ng20    &  \underline{\bf  74.37$\pm$0.16} &  73.80$\pm$0.17  &  64.12$\pm$0.20                  &  72.26$\pm$0.19                  &  63.07$\pm$0.21       &  72.61$\pm$0.20 \\  
  Ohscal  &  {\bf 58.95$\pm$0.19}            &  55.06$\pm$0.17  &  58.56$\pm$0.18                  &  58.66$\pm$0.21                  &  58.45$\pm$0.17       &  \underline{\bf 59.23$\pm$0.19} \\  
  R52     &  {\bf 83.87$\pm$0.24}  &  79.01$\pm$0.28  &  \underline{\bf 84.19$\pm$0.20}                  &  83.23$\pm$0.22                  &  83.36$\pm$0.21       &  {\bf 83.80$\pm$0.22} \\  
  R8      &     90.54$\pm$0.16               &  86.03$\pm$0.19  &  \underline{\bf 91.60$\pm$0.17}  &  90.24$\pm$0.17                  &  91.10$\pm$0.20       &  90.92$\pm$0.18 \\  
  Wap     &  16.47$\pm$0.34                  &  66.97$\pm$0.47  &  59.16$\pm$0.44                  &  \underline{\bf 70.18$\pm$0.54}  &  65.09$\pm$0.48       &  {\bf 70.02$\pm$0.53} \\  
  Webkb   &  73.29$\pm$0.39                  &  70.86$\pm$0.23  &  \underline{\bf 75.61$\pm$0.27}  &  74.19$\pm$0.37                  &  {\bf 75.59$\pm$0.29} &  74.97$\pm$0.35 \\ 
\hline\hline
\end{tabular}
\label{tbl_bin_mapl}
\end{table*}

\begin{table}[t]
\centering\small
\caption{Binary BoW representation: Win-loss-draw counts of measures in columns against those in rows based on the two standard error significance test over 10 runs.}
%\begin{tabular}{@{\hspace{4pt}} l @{\hspace{6pt}} | c @{\hspace{10pt}} c @{\hspace{10pt}} c @{\hspace{10pt}} c @{\hspace{10pt}} c @{\hspace{4pt}}}
\begin{tabular}{ l c c c c c }
\hline\hline
 &  {\it Sp} &  {\it WJac}  &  {\it WJac.idf}  & {\it Cos}  &  {\it Cos.idf}  \\
\hline\hline	       
{\it BM25}  &  5-2-3  &  5-5-0  &  4-3-2  &  4-4-2  &  3-7-0 \\
{\it Cos.idf}   &  8-1-1  &  5-4-1  &  8-1-1  &  4-6-0  \\	
{\it Cos}  &  7-2-1  &  5-3-2  &  6-3-1  \\		
{\it WJac.idf}  &  5-0-5  &  2-6-2 \\			
{\it WJac} &  8-0-2 \\				
\hline\hline
\end{tabular}
\label{tbl_bin_wld}
\end{table}

Table~\ref{tbl_bin_mapl} shows that {\it Sp} produced the best or competitive to the best result in six datasets; followed by {\it BM25} in five; {\it WJac.idf} in four; {\it Cos} in two; and {\it WJac} in one dataset only. {\it Cos.idf} did not produce competitive result to the best performing measure in any dataset. 

In terms of pairwise win-loss-draw counts as shown in the first column in Table~\ref{tbl_bin_wld}, {\it Sp} had many more wins than losses against all other contending measures.

It is interesting to note that {\it BM25}, {\it Cos.idf} and {\it Cos} using the binary BoW representation produced better retrieval results than their respective counterparts using the term-frequency-based BoW representation in some datasets. For example: (i) {\it BM25} in Fbis, New3s and Webkb; (ii) {\it Cos.idf} in La1s, Ohscal, Wap and Webkb; and (iii) {\it Cos} in La1s, La2s and Webkb. In contrast, {\it WJac.idf}, {\it WJac} and {\it Sp} using binary BoW vectors did not produce better retrieval results than their respective counterparts using term-frequency-based BoW vectors. 

Like in the term-frequency-based BoW representation, all measures had runtimes in the same order of magnitude.

\section{Discussion}
\label{sec_discussion}

Even though some studies have used different variants of tf and idf term weighting factors with the most widely used cosine similarity, the tf and idf factors discussed in Section~\ref{subsec_tf-idf} have been shown to be the most consistent in the IR task \citep{PhDThesis_Singhal1997}.

For the tf factor, instead of using the logarithmic scaling of $x_i$, some researchers have used other scaling approaches such as augmented $\left(0.5+0.5\times\frac{x_i}{\max(x_1,x_2,\cdots,x_M)}\right)$ \citep{TFIDF_Salton1988} and Okapi $\left(\frac{x_i}{2+x_i}\right)$ \citep{Okapi_Robertson1994}. Similarly, for the idf factor, instead of using $\frac{N}{n_i}$, some researchers have used the probabilistic idf factor based on $\frac{N-n_i}{n_i}$ \citep{Okapi_Robertson1994, PhDThesis_Singhal1997}. Note that BM25 (Eqn~\ref{eqn_bm25}) uses tf factor similar to Okapi and idf factor similar to the probabilistic idf factor \citep{ProbRelFramework_Robertson2009}.

In the supervised text mining task of document classification, different approaches utilizing class information are proposed to estimate the collection-based term weighting factors \citep{ICF_Wang2013, SupervisedTW_Debole2003, TermWeigthingTC_Lan2009}. Inverse category frequency (icf) \citep{ICF_Wang2013} has been shown to produce better classification result than the traditional idf factor with the cosine similarity measure. It considers the distribution of a term among classes rather than among documents in the given collection. The intuition behind icf is that the fewer classes a term $t_i$ occurs in, the more discriminating power the term $t_i$ contributes to classification \citep{ICF_Wang2013}. If $C$ and $c_i$ are the total number of classes and the number of classes in which $t_i$ occurs at least once in at least one document, then the icf factor is estimated as: $icf(t_i)=\log \left(1+\frac{C}{c_i}\right)$.

We have evaluated the performance of Sp in the kNN document classification task with existing measures using the supervised term weighting scheme of icf \citep{ICF_Wang2013}. Sp produced either better or competitive classification results with existing measures using supervised or unsupervised term weighting in the 5NN classification task. The classification results are provided in the Appendix.

Even though the weighted Jaccard similarity has been used in other application domains \citep{WeightedJacard_Chierichetti2010}, it is not widely used in the literature to measure similarities of BoW documents. Our experimental results in Section~\ref{sec_exp} show that the weighted Jaccard similarity with tf-idf term weighting scheme can be an effective alternative of cosine and BM25 in inter-document similarity measurements.

Sp has superior performance over all contenders in the binary BoW vector representation. It can be very useful in application domains such as legal and medical where the exact term frequency information may not available due to privacy issue because it is possible to infer information in a document from its term frequencies \citep{bigram_zhu2008}. 

\section{Concluding remarks}

For the purpose of inter-document similarity measurements task, we identify the limitations of the underlying assumptions of the most widely used tf-idf term weighting scheme employed in existing measures such as cosine and BM25; and provide an alternative which is more intuitive in this task. 

Based on the new assumption, we introduce a new simple but effective inter-document similarity measure called Sp. 

Our empirical evaluation in the query-by-example task shows that:

\begin{enumerate}
    \item Sp produces better or at least competitive results to the existing similarity measures with the state-of-the-art term weighting schemes in the term-frequency-based BoW representations. Sp produces more consistent results than the existing measures across different datasets.
    \item Sp produces better results than the existing similarity measures  with or without idf term weighting in the the case of binary BoW representation.
\end{enumerate}

When cosine and BM25 are employed, our result shows that it is important to use an appropriate BoW vector representation (binary or term-frequency-based) and also an appropriate term weighting scheme. Using inappropriate representation and term weighting scheme can result in poor performance. 

In contrast, using Sp, users do not have to worry about applying any additional term weighting to measure the similarity of two documents and still get better or competitive results in comparison to the best results obtained by cosine or BM25.

\section*{Acknowledgement}

The preliminary version of this paper was published in the Proceedings of the 11th Asia Information Retrieval Societies Conference 2015 \citep{tfidf_Aryal2015}.

% =======  Bibliography =====================
\bibliographystyle{abbrvnat}
\bibliography{SP}

% =======  Appendix =====================

\section*{Appendix A: kNN classification results}

In order to predict a class label for a test document ${\bf q}$, its $k$ nearest neighbour (or most similar) documents were searched in the given labelled training set of documents using a contending similarity measure and the majority class among the $k$NNs was predicted as the class label for ${\bf q}$. 

All classification experiments were conducted using a 10-fold cross validation (10 runs with each one out of the 10 folds as the test set and the remaining 9 folds as the training set). The average classification accuracy and standard error over a 10-fold cross validation were reported. All collection-based term weighting factors (idf and icf) were computed from the training set and used in both the training and test documents. The parameter $k$ was set to a commonly used value of 5 (i.e., 5NN classification was used).

We discuss the 5NN classification results with the term-frequency-based and binary BoW vector representations separately in the following two subsections.

\subsection*{{\bf Term-frequency-based BoW vector representation}}

We used term weighting based on tf only, tf-idf and tf-icf with weighted Jaccard and cosine resulting in eight contending measures: {\it Sp}, {\it BM25}, {\it Cos.tf-icf}, {\it Cos.tf-idf}, {\it Cos.tf}, {\it WJac.tf-icf}, {\it WJac.tf-idf} and {\it WJac.tf}. 

The average classification accuracies and standard errors over a 10-fold cross validation of the eight contending measures are provided in Table~\ref{tbl_tf_acc} and the summarized results in terms of pairwise win-loss-draw counts of contending measures based on the two standard errors significance test in the 10 datasets used in the experiment are provided in Table~\ref{tbl_tf_acc_wld}.

\begin{sidewaystable}
\centering\small
\caption{Term-frequency-based BoW representation: Average 5NN classification accuracy and standard error over a 10-fold cross validation. The best result is underlined and the results equivalent (insignificant difference based on two standard errors) to the best result are bold faced.}
\begin{tabular}{ l @{\hspace{20pt}} c @{\hspace{20pt}} c @{\hspace{20pt}} c @{\hspace{20pt}} c @{\hspace{20pt}} c @{\hspace{20pt}} c @{\hspace{20pt}} c @{\hspace{20pt}} c }
\hline\hline
  & {\it BM25.tf}   & {\it Cos.tf.icf}   & {\it Cos.tf.idf}   & {\it Cos.tf}   & {\it WJac.tf.icf}   & {\it WJac.tf.idf}   & {\it WJac.tf}   & {\it Sp} \\
  \hline\hline  
  Fbis  &   76.98$\pm$1.04   &   \underline{\bf 80.83$\pm$0.84}   &   {\bf 79.33$\pm$1.05}   &   {\bf 80.23$\pm$0.78}   &   {\bf 79.29$\pm$0.95}   &   {\bf 79.54$\pm$0.91}   &   {\bf 79.21$\pm$0.84}   &   {\bf 79.21$\pm$0.80} \\  
  La1s  &   83.68$\pm$0.58   &   83.43$\pm$0.84   &   86.70$\pm$0.66   &   82.30$\pm$0.80   &   87.30$\pm$0.80   &   \underline{\bf 88.89$\pm$0.61}   &   87.05$\pm$0.70   &   {\bf 88.48$\pm$0.46} \\   
  La2s   &   86.28$\pm$0.50   &   84.52$\pm$0.47   &   87.93$\pm$0.77   &   84.23$\pm$0.61   &   88.46$\pm$0.63   &   \underline{\bf 90.11$\pm$0.41}   &   88.03$\pm$0.56   &   {\bf 89.59$\pm$0.48} \\   
  New3s  &   79.31$\pm$0.30   &   79.54$\pm$0.34   &   78.99$\pm$0.34   &   \underline{\bf 81.47$\pm$0.29}   &   {\bf 80.90$\pm$0.30}   &   {\bf 80.86$\pm$0.33}   &   80.60$\pm$0.36   &   80.55$\pm$0.36 \\   
  Ng20  &   \underline{\bf 88.55$\pm$0.15}   &   87.57$\pm$0.22   &   86.92$\pm$0.22   &   84.74$\pm$0.34   &   86.28$\pm$0.27   & 87.41$\pm$0.25   &   83.05$\pm$0.28   &   86.62$\pm$0.17 \\   
  Ohscal  &   72.63$\pm$0.29   &   72.04$\pm$0.43   &   66.95$\pm$0.45   &   {\bf 74.25$\pm$0.46}   &   \underline{\bf 74.50$\pm$0.29}   &   72.36$\pm$0.21   &   {\bf 74.22$\pm$0.33}   &   73.19$\pm$0.34 \\   
  R52  &   \underline{\bf 92.30$\pm$0.20}   &   91.20$\pm$0.28   &   87.72$\pm$0.54   &   {\bf 92.18$\pm$0.18}   &   91.69$\pm$0.22   &   91.17$\pm$0.22   &   90.63$\pm$0.21   &   90.94$\pm$0.25 \\   
  R8  &   95.19$\pm$0.21   &   {\bf 95.34$\pm$0.17}   &   90.80$\pm$0.25   &   \underline{\bf 95.81$\pm$0.23}   &   {\bf 95.39$\pm$0.20}   &   94.98$\pm$0.23   &   {\bf 95.27$\pm$0.31}   &   95.28$\pm$0.27 \\
  Wap  &   17.76$\pm$0.79   &   75.90$\pm$0.46   &   76.92$\pm$0.76   &   72.44$\pm$0.58   &   80.70$\pm$0.80   &   {\bf 82.31$\pm$0.92}   &   76.22$\pm$0.58   &   \underline{\bf 82.50$\pm$0.79} \\   
  Webkb &   81.16$\pm$0.38 &   81.86$\pm$0.48 &   77.92$\pm$0.43 &   81.58$\pm$0.57 &   {\bf 84.14$\pm$0.42} &   83.33$\pm$0.61 &   \underline{\bf 84.40$\pm$0.38} &   {\bf 84.33$\pm$0.53} \\ 
\hline\hline
\end{tabular}
\label{tbl_tf_acc}
%\end{sidewaystable}
\vspace{0.8cm}
%\begin{sidewaystable}
\centering\small
\caption{Binary BoW representation: Average 5NN classification accuracy and standard error over a 10-fold cross validation. The best result is underlined and the results equivalent (insignificant difference based on two standard errors) to the best result are bold faced.}
\begin{tabular}{ l @{\hspace{20pt}} c @{\hspace{20pt}} c @{\hspace{20pt}} c @{\hspace{20pt}} c @{\hspace{20pt}} c @{\hspace{20pt}} c @{\hspace{20pt}} c @{\hspace{20pt}} c }
\hline\hline
  & {\it BM25}   & {\it Cos.icf}   & {\it Cos.idf}   & {\it Cos}   & {\it WJac.icf}   & {\it WJac.idf}   & {\it WJac}   & {\it Sp} \\
  \hline\hline
  Fbis  &   {\bf 79.29$\pm$0.65}   &   {\bf 79.62$\pm$0.99}   &   77.75$\pm$0.95   &   {\bf 78.20$\pm$0.88}   &   {\bf 79.82$\pm$1.08}   &   \underline{\bf 79.86$\pm$0.98}   &   {\bf 78.28$\pm$0.81}   &   {\bf 79.05$\pm$0.82} \\  
  La1s  &   84.27$\pm$0.63   &   {\bf 87.45$\pm$0.79}   &   {\bf 87.70$\pm$0.63}   &   85.89$\pm$0.72   &   {\bf 88.05$\pm$0.73}   &   \underline{\bf 88.70$\pm$0.54}   &   87.55$\pm$0.62   &   {\bf 88.67$\pm$0.49} \\  
  La2s   &   86.08$\pm$0.58   &   87.48$\pm$0.49   &   {\bf 89.43$\pm$0.42}   &   86.67$\pm$0.50   &   88.52$\pm$0.68   &   {\bf 89.99$\pm$0.43}   &   88.00$\pm$0.60   &   \underline{\bf 90.15$\pm$0.48} \\   
  New3s  &   \underline{\bf 80.72$\pm$0.42}   &   79.10$\pm$0.36   &   78.31$\pm$0.41   &   78.19$\pm$0.35   &   79.79$\pm$0.36   &   {\bf 80.03$\pm$0.41}   &   78.74$\pm$0.29   &   {\bf 80.15$\pm$0.40} \\   
  Ng20  &   \underline{\bf 87.59$\pm$0.13}   &   87.19$\pm$0.20   &   87.25$\pm$0.19   &   82.80$\pm$0.20   &   85.64$\pm$0.12   &   86.61$\pm$0.18   &   82.16$\pm$0.20   &   86.84$\pm$0.24 \\   
  Ohscal  &   72.02$\pm$0.32   &   72.36$\pm$0.31   &   68.54$\pm$0.32   &   72.89$\pm$0.24   &   \underline{\bf 73.65$\pm$0.28}   &   72.36$\pm$0.32   &   72.89$\pm$0.29   &   72.79$\pm$0.33 \\  
  R52  &   \underline{\bf 91.20$\pm$0.24}   &   89.74$\pm$0.38   &   86.14$\pm$0.43   &   90.21$\pm$0.27   &   90.51$\pm$0.25   &   89.81$\pm$0.24   &   89.75$\pm$0.17   &   {\bf 90.80$\pm$0.19} \\   
  R8  &   {\bf 94.80$\pm$0.21}   &   {\bf 95.05$\pm$0.13}   &   90.98$\pm$0.44   &   {\bf 94.99$\pm$0.23}   &   \underline{\bf 95.10$\pm$0.22}   &   94.54$\pm$0.30   &   {\bf 94.86$\pm$0.17}   &   {\bf 95.05$\pm$0.31} \\   
  Wap  &   15.51$\pm$0.69   &   76.86$\pm$0.65   &   78.27$\pm$1.01   &   69.68$\pm$0.82   &   80.00$\pm$0.56   &   \underline{\bf 81.92$\pm$0.96}   &   76.28$\pm$0.62   &   {\bf 81.60$\pm$0.81} \\   
  Webkb &   {\bf 83.97$\pm$0.49} &   {\bf 84.68$\pm$0.54} &   81.45$\pm$0.60 &   {\bf 84.26$\pm$0.49} &   \underline{\bf 84.88$\pm$0.44} &   {\bf 84.16$\pm$0.47} &   {\bf 84.73$\pm$0.42} &   {\bf 84.71$\pm$0.53} \\
\hline\hline
\end{tabular}
\label{tbl_bin_acc}
\end{sidewaystable}

\begin{table}[!htb]
\centering\small
\caption{Term-frequency-based BoW representation: Win-loss-draw counts of measures in columns against those in rows based on the two standard errors significance test over a 10-fold cross validation of 5NN classification.}
\begin{tabular}{l c c c c c c c}
\hline\hline
        & {\it Sp} &  {\it WJac.tf}  &  {\it WJac.tf-idf}  &  {\it WJac.tf-icf}  & {\it Cos .tf}  &  {\it Cos.tf-idf}  &  {\it Cos.tf-icf}  \\
\hline\hline
{\it BM25}  &  6-2-2  &  7-2-1  &  6-2-2  &  7-2-1  &  5-2-3  &  4-5-1  &  2-3-5  \\
{\it Cos.tf-icf}   &  6-1-3  &  5-2-3  &  5-0-5  &  6-1-2  &  4-2-4  &  2-5-3  \\
{\it Cos.tf-idf}   &  8-0-2  &  5-1-4  &  9-0-1  &  6-1-3  &  5-4-1  \\
{\it Cos.tf}  &  5-4-1  &  4-3-3  &  5-3-2  &  5-1-4  \\		
{\it WJac.tf-icf}  &  2-2-6  &  0-3-7  &  3-2-5 \\
{\it WJac.tf-idf}  &  1-1-8  &  2-5-3 \\
{\it WJac.tf} &  4-1-5 \\				
\hline\hline
\end{tabular}
\label{tbl_tf_acc_wld}
\end{table}

\begin{table}[!htb]
\centering\small
\caption{Binary BoW representation: Win-loss-draw counts of measures in columns against those in rows based on the two standard errors significance test over a 10-fold cross validation of 5NN classification.}
\begin{tabular}{l c c c c c c c }
\hline\hline
    & {\it Sp} &  {\it WJac}  &  {\it WJac.idf}  &  {\it WJac.icf}  & {\it Cos}  &  {\it Cos.idf}  &  {\it Cos.icf}  \\        
\hline\hline	       
{\it BM25}  &  4-1-5  &  4-3-3  &  3-2-5  &  4-3-3  &  3-3-4  &  3-6-1  &  3-3-4 \\
{\it Cos.icf}   &  4-0-6  &  0-1-8  &  3-2-5  &  3-1-6  &  0-4-6  &  1-5-4  \\	
{\it Cos.idf}  &  6-0-4  &  4-3-3  &  7-1-2  &  7-1-2  &  4-4-2  \\		
{\it Cos}  &  6-0-4  &  3-2-5  &  5-0-5  &  6-0-4 \\			
{\it WJac.icf}  &  3-1-6  &  0-5-5  &  3-3-4 \\
{\it WJac.idf}  &  1-0-9  &  0-4-6 \\
{\it WJac}  &  6-0-4 \\
\hline\hline
\end{tabular}
\label{tbl_bin_acc_wld}
\end{table}

The 5NN classification accuracies in Table~\ref{tbl_tf_mapl} show that {\it Sp}, {\it WJac.tf-idf}, {\it WJac.tf-icf} and {\it Cos.tf} produced the best or competitive to the best result in five datasets each followed by {\it WJac.tf} in four; {\it Cos.tf-icf} and {\it BM25} in two datasets each; and {\it Cos.tf-idf} in one dataset only. 

The pairwise win-loss-draw counts of {\it Sp} in the first column of Table~\ref{tbl_tf_acc_wld} shows that it had more wins than losses over all contending measures except {\it Wjac.tf-idf} and {\it Wjac.tf-icf} where it had competitive results with the same number of wins and losses. 

It is interesting to note that {\it Sp} and all three variants of weighted Jaccard similarity produced better classification results than all three variants of cosine and BM25. Like in the similar document retrieval task discussed in Section~\ref{sec_exp}, BM25 produced the worst classification accuracy in the Wap dataset because of $idf_{bm25}(t_i)$. The classification accuracy was increased to 79.42\% when $idf_{bm25}(t_i)$ was replaced by the traditional idf $idf(t_i)$.  

The supervised term weighting using icf (tf-icf) did not always produce better classification results than the traditional tf-idf based term weighting with both cosine and weighted Jaccard. It had five wins and two losses with cosine whereas it had two wins and three losses with weighted Jaccard. 

\subsection*{{\bf  Binary BoW vector representation}}

We used weighted Jaccard and cosine similarities with and without idf and icf weighting resulting in eight contending measures: {\it Sp}, {\it BM25}, {\it Cos.idf}, {\it Cos.icf}, {\it Cos}, {\it WJac.idf}, {\it WJac.icf} and {\it WJac}. 

The average classification accuracies and standard errors over a 10-fold cross validation of the eight contending measures are provided in Table~\ref{tbl_bin_acc} and the summarized results in terms of pairwise win-loss-draw counts of contending measures based on the two standard errors significance test in the 10 datasets used in the experiment are provided in Table~\ref{tbl_bin_acc_wld}.

The 5NN classification accuracies in Table~\ref{tbl_bin_acc} show that {\it Sp} produced the best or competitive to the best result in eight datasets. The closest contenders {\it BM25} and {\it WJac.idf} produced the best or competitive to the best result in six datasets each followed by {\it WJac.icf} in five; {\it Cos.icf} in four; {\it WJac} and {\it Cos} in three datasets each; and {\it Cos.idf} in two datasets only. 

In terms of pairwise win:loss:draw counts as shown in the first column in Table~\ref{tbl_bin_acc_wld}, {\it Sp} had more wins than losses against all other contending measures. It had one win and no loss against {\it WJac.idf} and three wins and one loss against {\it WJac.icf}. 

Like in the term-frequency-based BoW representation, the supervised weighting scheme based on icf did not always produce better classification results than the traditional idf based term weighting scheme with both cosine and weighted Jaccard in the binary BoW vector presentation as well. It had five wins and one loss with cosine whereas it had three wins and three losses with the weighted Jaccard. 

It is interesting to note that {\it BM25}, {\it Cos.icf}, {\it Cos.idf} and {\it Cos} which are using the binary BoW vector representation produced better classification accuracies than their respective counterparts using the term-frequency-based BoW representation in some datasets; e.g., {\it BM25} was better in three datasets (Fbis, New3s, WebKb); {\it Cos.icf} and {\it Cos} in three datasets (La1s, La2s, Webkb); and {\it Cos.idf} in two datasets (La2s, Webkb). However, all three variants of weighted Jaccard and {\it Sp} with the term-frequency-based BoW representation produced either better or competitive results with the binary BoW representation. 

\label{lastpage}

\end{document}